\documentclass[aps,shewpacs]{revtex4}
\usepackage{amsmath}
\usepackage{graphicx}
\usepackage{hyperref}
\usepackage{color}

\def\lsim{\mathrel{\raise3pt\hbox to 8pt{\raise -6pt\hbox{$\sim$}\hss{$<$}}}}
\def\rsim{\mathrel{\raise3pt\hbox to 8pt{\raise -6pt\hbox{$\sim$}\hss{$>$}}}}

\def\lsim{\mathrel{\raise3pt\hbox to 8pt{\raise -6pt\hbox{$\sim$}\hss{$<$}}}}

\newcommand{\boldtau}{\mbox{\boldmath $\tau$}}
\newcommand{\boldsigma}{\mbox{\boldmath $\sigma$}}

\renewcommand{\vec}[1]{{\mathbf #1}}

\newcommand{\bma}{\begin{pmatrix}}
\newcommand{\ema}{\end{pmatrix}}

\begin{document}

\draft
\title{Comparison between Variational Monte Carlo and Shell Model Calculations of Neutrinoless Double Beta Decay Matrix Elements in Light Nuclei}

\author{X.B. Wang$^a$, A.C. Hayes$^b$, J. Carlson$^b$, G.X. Dong$^a$,
E. Mereghetti$^b$, S. Pastore$^c$ and R.B. Wiringa$^d$}
\affiliation{
$^{\rm a}$\mbox{School of Science, Huzhou University, Huzhou 313000, China}
$^{\rm b}$\mbox{Theoretical Division, Los Alamos National Laboratory, Los Alamos, NM 87545, USA }
$^{\rm c}$\mbox{Department of Physics, Washington University in St. Louis, MO 63130, USA}
$^{\rm d}$\mbox{Physics Division, Argonne National Laboratory, Argonne, IL 60439, USA}
}

\date{\today}

\begin{abstract}
Benchmark comparisons between  many-body methods are performed to assess  the ingredients necessary for an accurate
calculation of neutrinoless double beta decay matrix elements.
Shell model and variational Monte Carlo (VMC) calculations are
carried out for $A=10$ and $12$ nuclei.
Different variational wavefunctions are used to evaluate the uncertainties in the {\it ab initio}
calculations, finding fairly small differences between the VMC double beta decay matrix elements.
For shell model calculations, the role of model space truncation, radial wavefunction choices,
and short-range correlation are investigated and all found to be important.
Based on the detailed comparisons between the VMC and shell model approaches,
we conclude that accurate descriptions of neutrinoless double beta decay matrix elements require
a proper treatment of both long-range and short-range correlations.
\end{abstract}
	
\maketitle

\section{Introduction}

Neutrinoless double beta decay ($0\nu\beta\beta$)  is a process in which
two neutrons in a nucleus decay into two protons, with the emission of two electrons and
no neutrinos, thus violating lepton number ($L$) by two units.
The observation of $0\nu\beta\beta$ would imply that neutrinos are Majorana
particles \cite{Schechter:1981bd},
shed light  on the  mechanism  of neutrino mass generation, and give insight into leptogenesis scenarios for the generation
of the matter-antimatter asymmetry in the universe \cite{Davidson:2008bu}. For these reasons,
$0\nu\beta\beta$ is the subject of intense experimental
research programs \cite{Gando:2012zm,Agostini:2013mzu,Albert:2014awa,Andringa:2015tza,KamLAND-Zen:2016pfg,Elliott:2016ble,Agostini:2017iyd,Aalseth:2017btx, Albert:2017owj,Alduino:2017ehq,Agostini:2018tnm,Azzolini:2018dyb}.
Current experimental limits are already very stringent,
constraining the half-lives of $^{76}$Ge, $^{130}$Te and $^{136}$Xe to be larger than  $8\cdot10^{25}$ yr
\cite{Agostini:2018tnm}, $1.5\cdot10^{25}$ yr \cite{Azzolini:2018dyb}
and $1.1\cdot10^{26}$ yr \cite{KamLAND-Zen:2016pfg}, respectively.
The next-generation ton-scale experiments will improve these limits by one or two orders of magnitude.

$0\nu\beta\beta$ rates depend not only on nuclear properties, but also on unknown
fundamental lepton number violating (LNV) parameters, such as the Majorana masses of light neutrinos.
Extracting the values of these parameters from experiments requires the evaluation of nuclear matrix elements (NMEs) of $0\nu\beta\beta$ transition operators.
Isotopes of experimental interest for 0$\nu\beta\beta$ searches, {\it e.g.} $^{48}$Ca, $^{76}$Ge, $^{82}$Se, $^{124}$Sn, $^{128}$Te,
$^{130}$Te and $^{136}$Xe, are medium and heavy open-shell nuclei with very complex nuclear structure. In addition, for these nuclei, one is forced by current computational limitations to utilize approximate methods to solve
the nuclear many-body problem, and to work in a truncated model spaces where correlations
and many-body terms in both the nuclear interactions and currents may be insufficient or neglected.
Consequently,  different theoretical models can give systematically different results.
For example, the nuclear matrix elements of $^{48}$Ca have been calculated using the shell model~\cite{Menendez:2008jp,Senkov:2013gso,Kwiatkowski:2013xeq,Iwata:2016cxn}, energy density functionals~\cite{Vaquero:2014dna}, the quasiparticle random-phase approximation (QRPA)~\cite{Simkovic:2013qiy}, and the interacting boson model (IBM)~\cite{Barea:2015kwa}, leading to results that differ by a factor of two or three.
Similar variations are observed for	 other $0\nu\beta\beta$ candidates \cite{Engel:2016xgb}.

Recently, a set of {\it ab initio} variational Monte Carlo (VMC) calculations
of 0$\nu\beta\beta$ decay matrix elements in $A=6-12$ nuclei has been reported
by some of the present authors \cite{Pastore:2017ofx}.  Within  the {\it ab initio}  VMC framework,
the many-body problem is solved  for nuclei up to $A=12$, with a nuclear
Hamiltonian consisting of two- and three-body interactions, namely the
Argonne-$v18$ (AV18) and Illinois-$7$ (IL7), respectively, and associated
electroweak many-body currents.
While the 0$\nu\beta\beta$  transitions  studied in
Ref.~\cite{Pastore:2017ofx} are not directly relevant
from an experimental point of view because the masses studied are too low,
they provide a useful reference for more approximated nuclear methods.
The {\it ab initio} framework  used in the VMC approach  accurately
explains,  qualitatively and  quantitatively,  the  observed electroweak
properties  of light nuclei \cite{Carlson:2014vla,Carlson:1997qn,Bacca:2014tla} over
a  broad range of momentum transfers, so that VMC results provide an important benchmark
to test other many-body methods that can be extended to the heavy nuclei of
experimental interest.

The goal of the present work is to benchmark
shell model calculations of the relevant $0\nu\beta\beta$ NMEs in light nuclei
to the aforementioned VMC results. We wish to examine the model dependence and
uncertainties involved in shell model approaches and to identify the degree of
sophistication that needs to be included in such a calculational approach to $0\nu\beta\beta$.
In general, shell model calculations involve a number of choices, including
the size of the model space, the effective nucleon-nucleon interaction,
the radial wave functions, and the short-range correlation (SRC) functions used.
In the current work, we concentrate on nuclei of mass 10 and 12, and we examine
the role played by these different nuclear structure inputs in
determining the predicted $0\nu\beta\beta$ NMEs.

The  paper  is  organized  as  follows. In  Section~\ref{SecII},
we present the two-body transition operators that mediate 0$\nu\beta\beta$,
and the shell model framework that is used to evaluate matrix elements of these operators.
In Section \ref{SecIII},  we  describe two sets of shell model calculations,
and present a detailed analysis of uncertainties arising from
different choices of the basis, radial functions, and SRC functions.
Finally,  in Section \ref{SecIV} we discuss our results and  present
our conclusions.

\section{Neutrinoless double beta decay matrix elements}\label{SecII}

\subsection{$0\nu\beta\beta$ operators}

We consider $0\nu\beta\beta$ transitions induced by a Majorana mass term for light,
left-handed neutrinos
\begin{equation}\label{eq:intro.0}
\mathcal L_{\Delta L = 2} = - \frac{m_{\beta\beta}}{2} \nu_{eL}^T C \, \nu_{eL},
\end{equation}
where $m_{\beta\beta} = \sum U_{e i}^2 m_{i}$, $m_i$ are the masses of the neutrino mass eigenstates,
and $U_{ei}$ are elements of the Pontecorvo-Maki-Nakagawa-Sato (PMNS) matrix. $C = i \gamma_2 \gamma_0$
denotes the charge conjugation matrix. The operator in Eq. \eqref{eq:intro.0} arises from the
Weinberg operator \cite{Weinberg:1979sa} after electroweak symmetry breaking, and the  $SU(2)_L$ invariance
of the Standard Model implies that $m_i \sim v^2/\Lambda$, where $v$ is the Higgs vacuum expectation value
and $\Lambda$ is the high-energy scale at which LNV arises.
Eq. \eqref{eq:intro.0} is the first term in an expansion in $v/\Lambda$, and, in general,
$0\nu\beta\beta$ will receive contributions from operators of dimension
higher than five \cite{Pas:2000vn,Pas:1999fc,Prezeau:2003xn,Lehman:2014jma,Graesser:2016bpz,Cirigliano:2017djv,Cirigliano:2018yza}.
We limit ourselves to study nuclear matrix elements of the $0\nu\beta\beta$ transition operator induced by  $m_{\beta\beta}$,
since its structure is rich enough to capture several features of non-standard mechanisms.

The $0\nu\beta\beta$ transition operator, or ``neutrino potential", induced by
$m_{\beta\beta}$ has been derived in several papers \cite{Haxton:1985am,Doi:1985dx,Bilenky:2014uka}.
In general, several operators $M_\alpha^\beta$ contribute, with $\alpha \in \{ F, GT, T\} $,
and  $\beta \in \{\nu, AA, AP, PP, MM\}$ labeling the various components of the Fermi(F),
Gamow-Teller(GT) and tensor(T) matrix elements of the vector ($\nu$), axial-axial(AA),
axial-pseudoscalar (AP), pseudoscalar-pseudoscalar (PP), and magnetic-magnetic (MM) operators.
The leading operators are,
\begin{eqnarray}
\label{oalpha}
\nonumber M^{\beta}_{GT} &=& (4\pi R_A)\vec{\sigma}_1 \cdot \vec{\sigma}_2 V^{\beta}_{GT}(r_{12}) \tau^+_1\tau^+_2\ , \\
\nonumber M^{\beta}_{F} &=& (4\pi R_A) V^{\beta}_{F}(r_{12})\ ,  \\
M^{\beta}_{T} &=&(4\pi R_A) \left[3\left(\vec{\sigma}_1 \cdot \hat{r} \right)
\left(\vec{\sigma}_2 \cdot \hat{r} \right) -
\vec{\sigma}_1 \cdot \vec{\sigma}_2 \right] V^{\beta}_{T}(r_{12})\ ,
\end{eqnarray}
where $R_A= 1.2A^{1/3}$ fm is the nuclear radius.
The form of the neutrino potentials $V^{\beta}_{\alpha}(r_{12})$ in coordinate space is given
in Ref.~\cite{Pastore:2017ofx}, where the corresponding potential in momentum space is also listed.
The  functions $V_\alpha^{\beta}$ in Eq. \eqref{oalpha} have different radial dependencies.
In particular, $V_F^\nu$ and $V^{AA}_{GT}$, which give the largest contributions to the NMEs,
have a Coulombic $1/r$ dependence, up to corrections from the axial and vector form factors.
The $AP$ and $PP$ components are induced by the induced pseudoscalar form factor, which is
dominated by the pion pole, and thus have pion range.
The $MM$ components are generated by neutrinos coupling to the weak magnetic form factor, and have short-range.
Finally, it was pointed out in Ref.~\cite{Cirigliano:2018hja} that, in chiral effective field theory ($\chi$EFT),
the $0\nu\beta\beta$ transition operator should be supplemented by a leading-order short-range component,
in addition to $MM$, with unknown coefficient.






\subsection{Variational Monte Carlo Method}
\label{sec:vmc}

In this work, VMC calculations used the same wave functions as
constructed in Ref.~\cite{Pastore:2017ofx}. Here, we summarize
the calculations scheme adopted in that reference.

The evaluation of the matrix elements defined in
Eq.~(\ref{oalpha}) is carried out using VMC computational
algorithms~\cite{Carlson:2014vla}.
The VMC wave function $\Psi(J^\pi;T,T_z)$---where $J^\pi$
and $T$ are the spin-parity and isospin of the state---is
constructed from products of two- and three-body correlation
operators acting on an antisymmetric single-particle state of
the appropriate quantum numbers.
The correlation operators are designed to reflect the influence of the
two- and three-body nuclear interactions at short distances, while
appropriate boundary conditions are imposed at long range~\cite{Wiringa:1991kp,Pudliner:1997ck}.

The $\Psi(J^\pi;T,T_z)$ has embedded variational parameters
that are adjusted to minimize the expectation value
\begin{equation}
 E_V = \frac{\langle \Psi | H | \Psi \rangle}
            {\langle \Psi   |   \Psi \rangle} \geq E_0 \ ,
\label{eq:expect}
\end{equation}
which is evaluated by Metropolis Monte Carlo integration~\cite{Metropolis:1953am}.
In the equation above, $E_0$ is the exact lowest eigenvalue of the nuclear
Hamiltonian $H$ for the specified quantum numbers.
The many-body Hamiltonian is given by
\begin{equation}
 \label{eq:nucH}
 H = \sum_{i} K_i + {\sum_{i<j}} v_{ij} + \sum_{i<j<k} V_{ijk} \ ,
\end{equation}
where $K_i$ is the non-relativistic kinetic energy of nucleon $i$
and $v_{ij}$ and $V_{ijk}$ are, respectively,
the Argonne $v_{18}$ (AV18)~\cite{Wiringa:1994wb}
two-body potential and the Illinois-7
(IL7)~\cite{Pieper:2008rui} three-nucleon interaction.
The AV18+IL7 model reproduces the experimental
binding energies, charge radii, electroweak transitions
and responses of $A=3$--$12$ systems in numerically exact
calculations based on Green's function Monte Carlo (GFMC)
methods~\cite{Carlson:1997qn,Bacca:2014tla,Carlson:2014vla,Pastore:2017uwc}.

A good variational wave function, that serves as the starting
point of GFMC calculations, can be constructed with
\begin{equation}
   |\Psi_V\rangle =
      {\cal S} \prod_{i<j}^A
      \left[1 + U_{ij} + \sum_{k\neq i,j}^{A}\tilde{U}_{ijk} \right]
      |\Psi_J\rangle.
\label{eq:psit}
\end{equation}
The Jastrow wave function $\Psi_J$ is fully antisymmetric, translationally
invariant, has the
$(J^\pi;T,T_z)$ quantum numbers of the state of interest, and includes
a product over pairs of a central correlation function $f(r_{ij})$
that is small at short distances, peaks around 1 fm, and decays
exponentially at long range~\cite{Pieper:2001mp}.
The $U_{ij}$ and $\tilde{U}_{ijk}$ are the two- and three-body correlation
operators, and $\cal S$ is a symmetrization operator.
The two-body correlation operators~\cite{Pieper:2001mp,Carlson:2014vla} can
be schematically written as
\begin{equation}
 U_{ij} = \sum_p f^p(r_{ij}) \, O^p_{ij} \ ,
\end{equation}
where
\begin{equation}
O^p_{ij}= \boldtau_i\cdot \boldtau_j\,,\, \boldsigma_i\cdot\boldsigma_j\,,\,
(\boldtau_i\cdot\boldtau_j)(\boldsigma_i\cdot\boldsigma_j)\,,\,S_{ij}\,
,\,S_{ij}\boldtau_i\cdot \boldtau_j\ ,
\end{equation}
are the main static operators that appear in the two-nucleon potential
and the $f^p$ are functions of the interparticle distance $r_{ij}$
generated by the solution of a set of coupled differential equations
containing the bare two-nucleon potential with asymptotically-confined
boundary conditions~\cite{Carlson:2014vla}.

The results presented below for $A=10$ nuclei use the VMC wave functions that
serve as starting trial functions for the GFMC calculations in Ref.~\cite{Carlson:2014vla}.
For $A=12$ systems, we use both shell-model-like wave functions (denoted
in the figures by ``VMC-1") and clusterized wave functions (denoted by ``VMC-2")
which were originally used in the calculations reported in Ref.~\cite{Nollett:2000ch}.
The difference in these two types of variational wave functions is described
below.

For $A > 4$ nuclei, the Jastrow wave function $\Psi_J$ of Eq.~(\ref{eq:psit}) is written as
a sum over different $LS$-space symmetry combinations $^{2S+1}L_J[n]$
(where $[n]$ denotes the spatial symmetry Young diagram),
each of which is an antisymmetric sum over partitions of $A$ nucleons into
four s-shell nucleons and $A-4$ p-shell nucleons.
For the shell-model-like wave function, each partition has a product over all
pairs of a set of three central correlation functions, $f_{xy}(r_{ij})$,
where $xy = ss$ indicates both nucleons are in the s-shell, $xy = pp$ indicates
both are in the p-shell, and $xy = sp$ that one is in each shell.
Each of the $f_{xy}$ has a different radial dependence with variational
parameters to be optimized, but all p-shell particles are treated equally.
Thus for a 12-body nucleus, there is a product of 6 $f_{ss}$, 32 $f_{sp}$,
and 28 $f_{pp}$ functions.
In addition, the parametrization of the $f_{xy}$ are allowed to be different
for different $^{2S+1}L_J[n]$ components.
For the $^{12}$C ground state we use $^1$S$_0$[444] and $^3$P$_0$[4431] spatial
symmetry combinations, with coefficients determined in a $2 \times 2$ energy
diagonalization.
For $^{12}$Be we use $^1$S$_0$[4422] and $^3$P$_0$[4332] combinations.

For the cluster-type wave functions, we allow for the formation of clusters
in the p-shell by further partitioning the $A-4$ nucleons into subgroups and
having multiple types of $f_{pp}$ and $f_{sp}$.
For the $^{12}$C $^1$S$_0$[444] ground state there are two $f_{pp}$ correlation
functions, used according to whether both nucleons are in the same or different
[4] clusters; effectively this builds a triple-alpha structure which is
a major part of the $^{12}$C ground state.
For the $^{12}$Be $^1$S$_0$[4422] ground state there are three subgroups in the
p-shell, effectively one alpha-like [4] cluster and two dineutron-like [2]
clusters, with three $f_{pp}$ and two $f_{sp}$ correlation functions.
These cluster-type wave functions are more sophisticated in construction,
but the increase in the number of variational parameters to optimize is a
burden, and only these highest spatial symmetry states have been used so far.

At present, the shell-model-like VMC-1 wave function for $^{12}$C with two
spatial symmetry components gives a slightly better energy, while getting a
charge radius closer to experiment than the cluster-like VMC-2 wave function
with one spatial symmetry.
The VMC-1 and VMC-2 wave functions for $^{12}$Be give comparable charge radii
close to experiment, but the VMC-2 allows the neutron distribution to be more
diffuse with a slightly better energy.

In what follows, we calculate matrix elements defined
in Eq.~(\ref{oalpha}), and their associated
transition distributions in $r$-space,
$C^{\alpha,\beta}(r)$, and $q$-space, $\bar{C}^{\alpha,\beta}(q)$
defined as
 \begin{eqnarray}
\label{eq:densities}
 M^{\alpha}_\beta &=& \int d{\bf r}\, \rho^{\alpha}_\beta(r)
 \equiv \int dr \, C^{\alpha}_\beta(r) \equiv \int dq \, \bar{C}^{\alpha}_\beta(q) \ , \nonumber\\
\end{eqnarray}
where  $\rho^{\alpha}_\beta(r)$ is the transition density associated
with the transition operator $O^{\alpha}_\beta(r)$.

\subsection{NMEs within the shell model framework}

Within the shell model, the matrix elements of the $M^\alpha_\beta$ operators
between many-body wave functions $\mid i >$ and $\mid f>$
can be calculated as a sum of products
of two-body transition densities (TBTDs) between many-body states and antisymmetrized two-body matrix elements TBMEs between two-particle states, as
\begin{equation}
\label{nme0}
\langle f|M^\alpha_\beta| i\rangle = \sum_{J_0,k_{\alpha}\leq  k_{\beta},k_{\gamma} \leq k_{\delta}}\texttt{TBTD}(f,i,k,J_0)\langle k_{\alpha}k_{\beta},J_0  | M^\alpha_\beta| k_{\gamma}k_{\delta},J_0 \rangle \;.
\end{equation}
Here the labels $k$ stands for the set of spherical quantum numbers $(n,l,j)$ describing the single nucleons making up the two-particle wave functions,
and $\texttt{TBTD}$ are matrix elements between the many-body wave functions given by,
\begin{equation}
\label{tbtd}
\texttt{TBTD}(f,i,k,J_0)= \langle f || [A^+(k_{\alpha}k_{\beta},J_0) \otimes \tilde{A}( k_{\gamma}k_{\delta},J_0)  ]^{(0)} || i\rangle,
 \end{equation}
where $A^+$ is a two-particle creation operator of rank $J_0$, as
\begin{equation}
\label{a+}
A^+(k_{\alpha}k_{\beta},J_0, M_0) = \frac{[a^+(k_{\alpha})a^+(k_{\beta})]^{J_0}_{M_0}}{\sqrt{(1+\delta_{k_{\alpha}k_{\beta}})}},
 \end{equation}
and $ \tilde{A}(k_{\alpha}k_{\beta},J_0) = (-1)^{J_0-M_0} A^{\dagger}(k_{\alpha}k_{\beta},J_0, M_0) $.

\section{Calculations and discussions}\label{SecIII}

We focus on the $0\nu\beta\beta$ NMEs for
the $\Delta T=0$ $^{10}$Be$\rightarrow^{10}$C and
$\Delta T=2$ $^{12}$Be$\rightarrow ^{12}$C transitions.
The $\triangle T=0$ transition is between isobaric analog states,
a situation that is  never realized in $0\nu\beta\beta$ experiments,
but is nonetheless
very useful for comparisons between models. In particular, as discussed below,
the radial contributions to the A=10 $0\nu\beta\beta$ matrix elements involve no nodes,
in contrast to the A=12 case. As in Eq.~(\ref{eq:densities}), the radial
contribution to the matrix elements is defined as $C(r)$, where $\int_0^\infty C(r) dr = M_\alpha^\beta$.

Our shell model calculations are carried out using the code  {\sc Oxbash}~\cite{oxbash}.
The PSDMWK shell model Hamiltonian~\cite{Warburton:1992rh,Warburton:1992qu}
is used for the p- plus sd-shell ($psd$) model space, with
up to $4p4h$ excitations.  Lawson's prescription is adopted for treating
center of mass (C.M.) excitations ~\cite{Gloeckner:1974sst}.
The same NMEs are also calculated by VMC, for which the
Argonne $v_{18}$  two-nucleon potential plus Illinois-7 three-nucleon interaction are used.
As discussed in Sec.~\ref{SecII}, two different sets of variational wave functions
are used in the VMC calculations, namely the shell-model-like (VMC-1) and cluster-model-like (VMC-2).

\subsection{The wave function normalization}

\begin{figure}
\includegraphics[angle=0,width=12 cm]{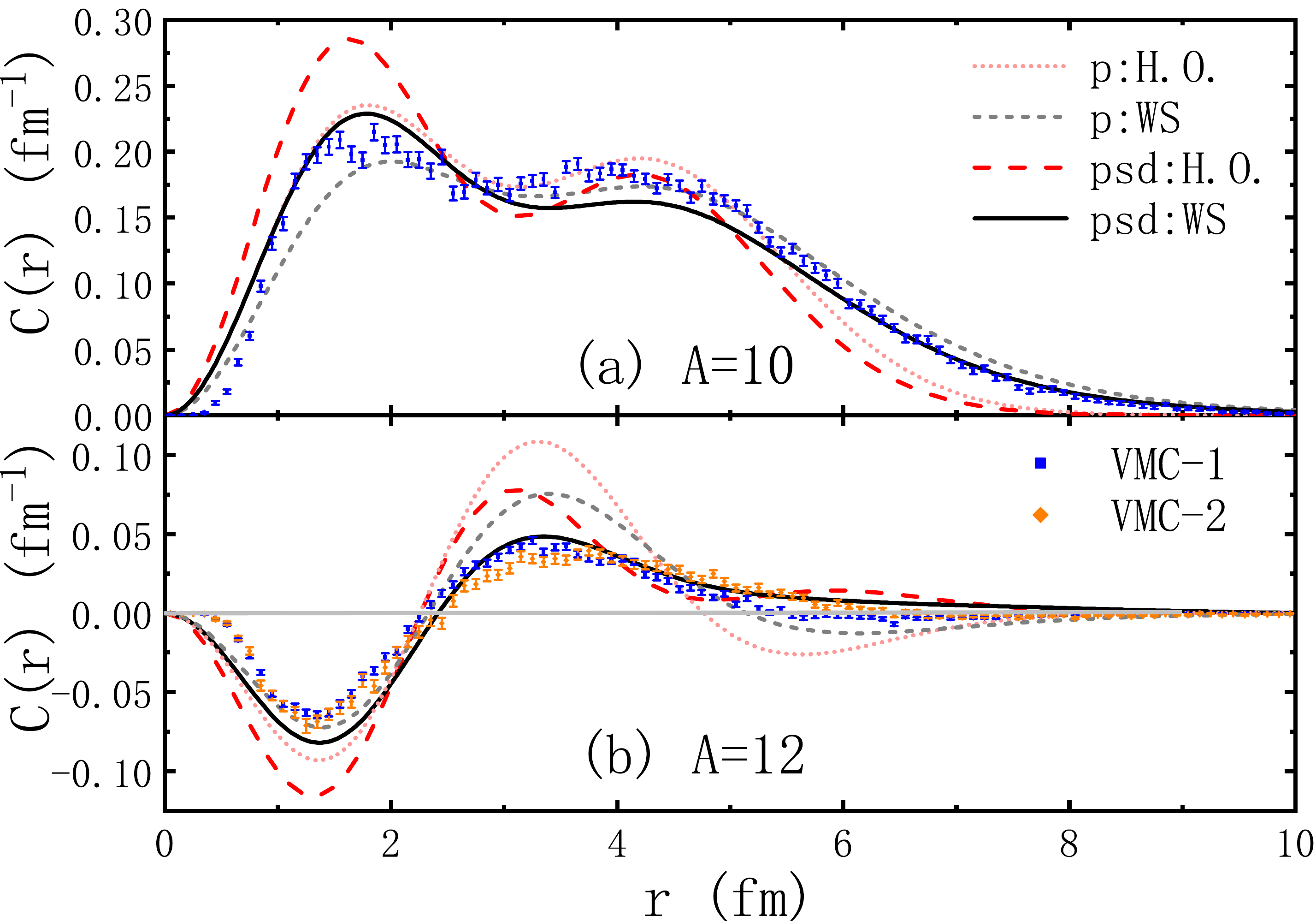}
\caption[T]{\label{fig:norm}
(Color online)
The normalization densities $C(r)=\langle f|\sum_{a<b}\delta(r-r_{ab})\tau_a^+ \tau_b^+| i\rangle$, for (a) A=10, and (b) A=12.
The $0\hbar\omega$ model space ($p$-shell only) and extended model space of $psd$ shell model calculation results are shown.
The different choices of radial wave functions, H.O., and WS, are also shown.
VMC results with shell-model-like wave functions are labeled as ``VMC-1", and
those with cluster-like wave functions are labeled as ``VMC-2".
}
\end{figure}

We begin our study with a comparison of the wave function normalization densities
\begin{equation}\label{Cr}
C(r)=\langle f|\sum_{a<b}\delta(r-r_{ab})\tau_a^+ \tau_b^+| i\rangle,
\end{equation}
displayed in Fig.~\ref{fig:norm}. For the shell model calculations, we show the results
for the $p$ shell and the  more extended  $psd$ model space. In addition, we show shell
model results using different choices of radial wave functions.
The SRC functions are not included at this stage in the shell model wave functions.

Harmonic oscillator (H.O.) radial wave functions are commonly used in shell model
calculations because H.O. basis have the advantage of a simple well-defined method
of separating into center-of-mass and relative coordinates via Talmi-Brody-Moshinsky
brackets~\cite{Talmi52,MOSHINSKY59}. However,  H.O. wave functions exhibit the wrong asymptotic behavior
at large interparticle distance, $r$, which affects the calculation of $0\nu\beta\beta$ NMEs.
Thus, we examine the effect of using more realistic Woods-Saxon (WS) radial wave functions.
In the case of WS wave functions, we take the $p$-shell neutron and proton asymptotic wave
function behavior to be determined by the corresponding experimental separation energies,
while the unbound $sd$-shell particles are assumed to be bound by $0.05$ MeV.
For the calculations of two-body matrix elements, we expand the WS wave functions
into a H.O. basis, and then apply the Talmi-Moshinsky method.

As seen in both panels of Fig.~\ref{fig:norm}, the extended shell model space results in a higher
first (and lower second) peak in $C(r)$.
Enlarging the model space introduces more correlations, making the nuclei more bound, and the wave
functions more concentrated at short distances.
The more realistic WS wave functions provide a better description at large $r$.
We  note that the normalization of $C(r)$ is unity for the A=10 $\triangle T=0$ isobaric analog
transition, and zero for $\Delta T=2$ A=12 transition.
The position of the nodes in the $\Delta T=2$ $C(r)$ function
has an important effect on the predictions for the $0\nu\beta\beta$ NMEs.

As seen in panel (b) of Fig.~\ref{fig:norm}, the first node appears at about 2 fm in
both the shell model and VMC calculations.
However, the smaller $p$-shell model space has an extra node at
around 5 fm, in contrast to the predictions of the
larger model spaces.
Neither VMC calculations predict this second node.
Thus, we conclude that the size of the model space used is
crucial for shell model estimates of $0\nu\beta\beta$ NMEs.

\subsection{The radial distribution of NMEs}
\begin{figure}
\includegraphics[angle=0,width=18 cm]{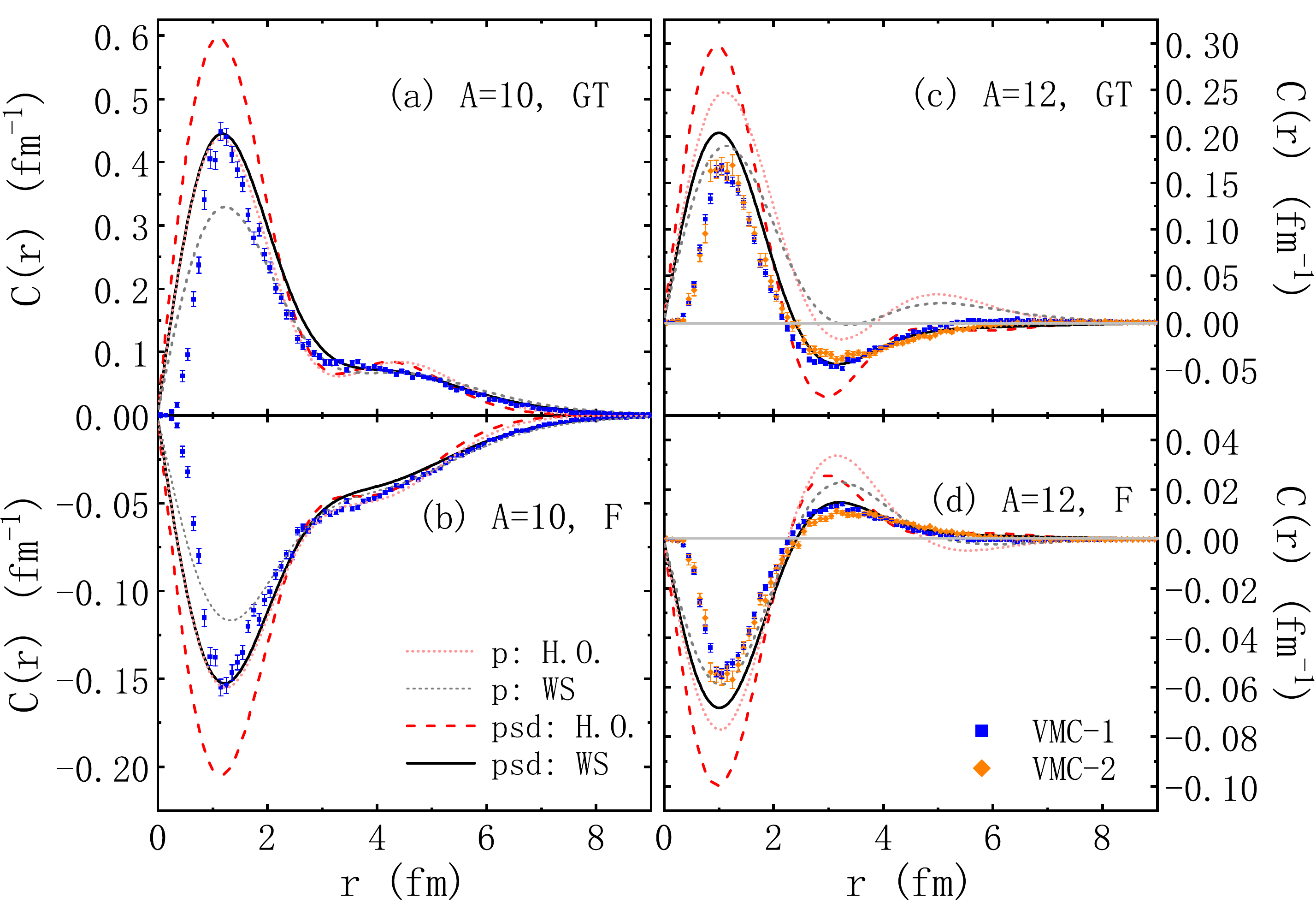}
\caption[T]{\label{fig:rnme}
(Color online)
The radial distributions associated with F-${\nu}$ and GT-AA  operators, for (a)-(b) A=10, and (c)-(d) A=12 nuclei.
F-${\nu}$ and GT-AA matrix element contributions are labeled as ``F" and ``GT ", respectively.
The plots show the radial distributions $C(r)$ determining the $0\nu\beta\beta$ matrix elements,
where $\int_0^\infty C(r) dr =M_\alpha^\beta$.
The shell model results are for the lowest order
$0\hbar\omega$ $p$-shell calculations and $psd$-shell calculations with up
to four particle excitations, and for two choices of radial wave functions, H.O. and WS, without short-range correlations.
VMC results are the full calculations and include statistic error bars.
VMC results with shell-model-like wave functions are labeled as ``VMC-1", and those with cluster-like wave functions are labeled as ``VMC-2". The
radius factor $R_A$ of Eq.~(\ref{oalpha}) is not included in the curves.
}
\end{figure}

We next consider the matrix elements of $V_F^\nu$ and $V^{AA}_{GT}$, which have a simple Coulombic $1/r$ dependence and give the largest contributions to the NMEs.
Detailed comparisons between the shell model and VMC calculations are shown in Fig.~\ref{fig:rnme}.
Apart from the differences in magnitude and sign, the distribution $C(r)$ for $V_F^\nu$ and $V^{AA}_{GT}$ have similar shapes.
Because of the Coulombic $1/r$ dependence, the first peak of the distributions is much larger than the second peak,
and the functions die off rapidly at large distance $r$.
As with the normalization functions, the shell model distributions from the larger $psd$ model spaces
show higher peaks than those from the smaller $p$-shell model spaces, when the same radial wave functions are used.
In general, the predicted shape of the functions $C(r)$ is poor in the case of the small pure $p$-shell calculations,
and we again conclude that small shell model spaces are unlikely to provide accurate $0\nu\beta\beta$ predictions.
When a larger model space is used in combination with the  more realistic WS wave functions, the shell model predictions
are in reasonable agreement with the VMC results, except at very small $r<1$ fm.

\subsection{The short-range correlation functions}

\begin{figure}
\includegraphics[angle=0,width=10 cm]{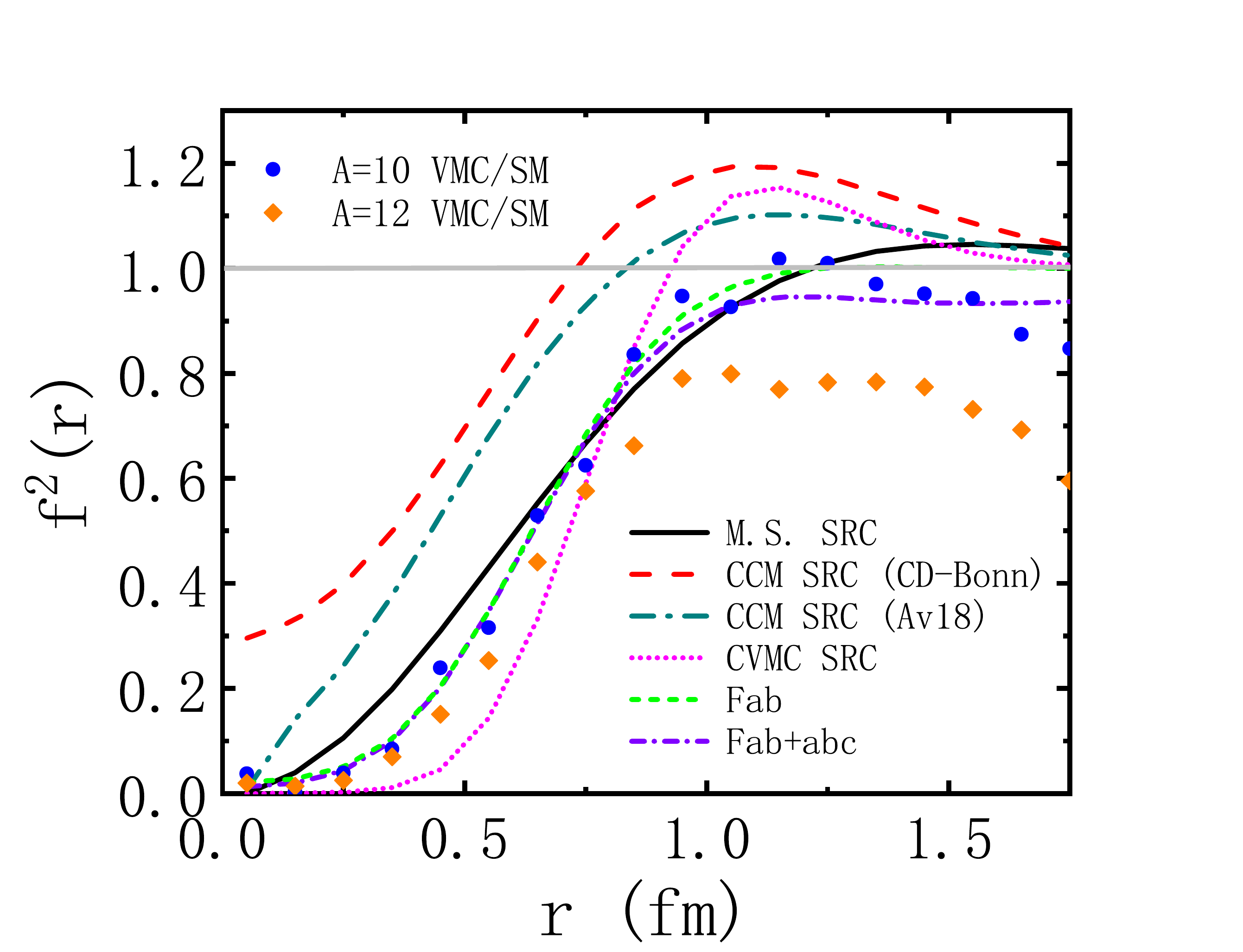}
\caption[T]{\label{fig:fr}
(Color online)
The square of the Jastrow-like correlation functions (Eq.~(\ref{fsrc})). The different correlation functions are
described in the text. The points correspond to the ratio of the VMC to shell-model (with $psd$ model space and WS radial wave functions) normalization functions for A=10 and 12.}
\end{figure}

Short-range correlations (SRC) arising from the repulsive hard core of the nucleon-nucleon interaction are absent from the radial wave functions used in shell model calculations.
A standard approximate method for correcting this is to introduce a
Jastrow-like correlation function, $f(r\equiv\mid r_1-r_2\mid)$, of the form,
\begin{equation}
\label{fsrc}
f(r) =1 -c e^{-ar^2} \left( 1-br^2 \right) \ .
\end{equation}
The parameters $a,b,c$ have been determined using different assumptions.
The Miller-Spencer (M.S.) SRC, modeled in 1976~\cite{MILLER1976562}, is widely used in the literature.
Newer sets of SRC have been obtained from $^1S_0(n=0)$ correlated two-body wave function derived by
the coupled-cluster method (CCM)~\cite{Simkovic:2009pp}.
More recently, through the study of paired density distributions in $^{16}$O and $^{40}$Ca by
Cluster Variational Monte Carlo (CVMC)~\cite{CVMC1992,LonardoniCVMC}, isospin dependent correlation
functions have been suggested~\cite{CRUZTORRES2018304}.
Yet another set of  correlation functions was obtained in Ref.~\cite{Pandharipande1997} from  nuclear
matter variational calculations, and we also study the effects of using these functions.
In Figs.~\ref{fig:fr} and~\ref{fig:rnmesrc},
two-body correlation functions from Ref.~\cite{Pandharipande1997} with and without three-body
and higher order correlations are labeled with ``Fab+abc" and ``Fab", respectively.

In Fig.~\ref{fig:fr}, we show a comparison between the different Jastrow correlation functions
which we used in calculating the shell model NMEs. This allows us to assess
the sensitivity  of our shell model results with respect to variations in the SRCs.
The points in Fig.~\ref{fig:fr} denote the ratio of the normalization density
$C(r)$, defined in Eq.~\eqref{Cr}, computed in the VMC and in the shell-model for A=10 and 12.
This ratio might be regarded as a rough estimate of short-range dynamics present in VMC
versus shell model calculations.

As seen in the figure, most of short-range correlations overshoot unit at intermediate distance, and exhibit a correlation ``hole" at short distance. Such overshoot is needed to preserve the wave function normalizations, as has been studied in Ref.~\cite{Jastrow2011}.
The Miller-Spencer SRC function peaks at about 1.5 fm, while more modern functions, CCM SRC and CVMC SRC,
peak at 1.0 fm. The CVMC SRC exhibits a more pronounced peak and hole relative to the CCM SRC (A$v18$). As discussed in the next section, the difference in NME predictions obtained using the different SRC functions reflects a general uncertainty in the shell model calculations.



\subsection{Results}
\begin{figure}
\includegraphics[angle=0,width=18 cm]{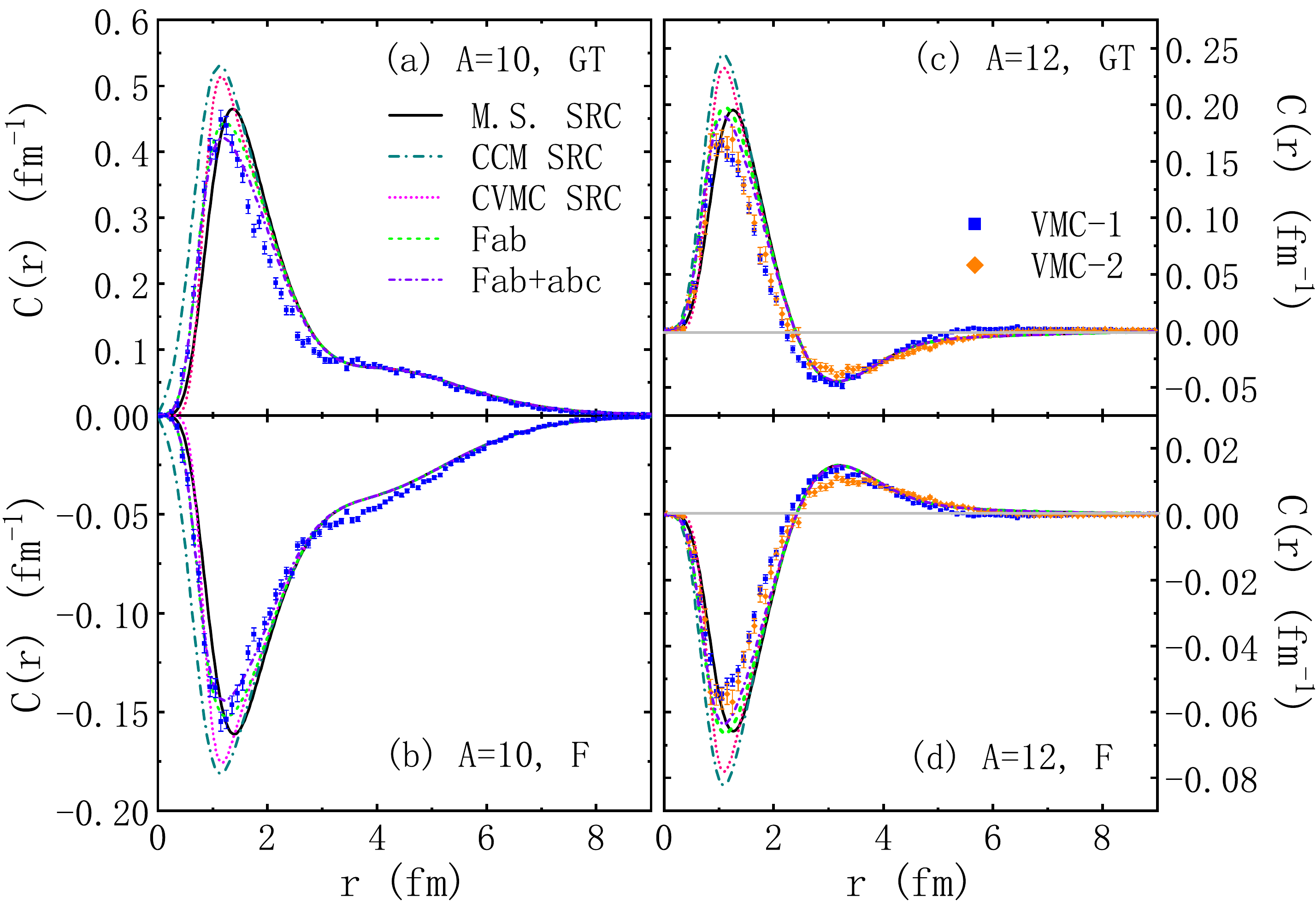}
\caption[T]{\label{fig:rnmesrc}
(Color online) The same as in Fig.~\ref{fig:rnme}, but for the shell model calculations of NMEs involving SRC.
}
\end{figure}

In Fig.~\ref{fig:rnmesrc}, shell model matrix elements using the different SRC functions are shown.
In these calculations, $psd$ shells with WS wave functions are used.
As expected, the degree of suppression at small $r$ of the function $C(r)$ is dependent on the  SRC used.
The resulting  NMEs are given in Table.~\ref{nme}.
The VMC calculations with the two different type of wave functions give very similar results.
However, within a  shell model framework, several important ingredients must be included
for realistic predictions.
These include the size of the model space, the choice of radial wave functions and the inclusion
of realistic SRC functions.

For $A$=10, the extension of shell model space results in larger (in absolute value) NMEs.
For $A$=12, the extension of the model space results in smaller  matrix elements, because of the cancellations from contributions above and below the node in $C(r)$.
The use of more realistic radial wave functions, WS versus HO, also reduces the predicted NMEs.
The inclusion of a SRC function is important to further reduce the shell model predictions and to obtain realistic functions $C(r)$.
In Ref.~\cite{KORTELAINEN2007128}, it was reported that for $^{48}$Ca and $^{76}$Ge, a 30 to 40 $\%$ reduction of NMEs arose from the inclusion of the M.S. SRC. In the present study of light nuclei, the M.S. SRC reduces the NMEs by about 20 $\%$. Among more modern SRC functions, respecting the preservation of wave-function normalization~\cite{Jastrow2011}, SRC fitted from CVMC also gives similar reduction of NMEs.

\begin{table}
	\caption{Results of dimensionless matrix elements F-${\nu}$ and GT-AA matrix elements,  labeled as ``F" and ``GT ".
VMC results with different variational wave functions are labeled as ``VMC-1" and ``VMC-2". VMC statistical errors are also listed. Shell model results with different radial wave functions (H.O. and WS), model space ($p$ and $psd$) and SRC are given correspondingly.The radius factor $R_A$ from eqs. (2)  is included.}
	\begin{center}
        \begin{tabular}{c|cc|cc}
        \hline
     \hline
       & \multicolumn{2}{c}{$^{10}$Be($0^+_1$)$\rightarrow^{10}$C($0^+_1$)}   &\multicolumn{2}{c}{$^{12}$Be($0^+_1$)$\rightarrow^{12}$C($0^+_1$)}\\
       &  \multicolumn{1}{c}{F} &\multicolumn{1}{c}{GT} &\multicolumn{1}{c}{F} &\multicolumn{1}{c}{GT} \\
        \cline{1-5}
\vspace{-0.1in}\\
     VMC-1&-1.001(40)&2.273(91)&-0.100(4)&0.257(10)\\
     VMC-2& ---&---&-0.113(5)&0.274(11)\\\hline
     SM$_{\texttt{H.O.}}$(w/o SRC,~$p$)&-1.127&2.616&-0.183&1.228\\
      SM$_{\texttt{WS}}$(w/o SRC,~$p$) & -0.980&2.269&-0.147&1.023\\
      SM$_{\texttt{H.O.}}$(w/o SRC,~$psd$)&-1.274&3.228&-0.271&0.431\\
      SM$_{\texttt{WS}}$(w/o SRC,~$psd$) & -1.100&2.783&-0.198&0.570\\\hline
     SM$_{\texttt{WS}}$(M.S. SRC,~$psd$)  &-0.967&2.381&-0.122&0.342\\
     SM$_{\texttt{WS}}$(CCM SRC,~$psd$)  &-1.069&2.683&-0.175&0.499\\
     SM$_{\texttt{WS}}$(CVMC SRC,~$psd$)  &-0.992&2.457&-0.141&0.398\\
     SM$_{\texttt{WS}}$(Fab,~$psd$) &-0.988&2.449&-0.138&0.388\\
     SM$_{\texttt{WS}}$(Fab+abc,~$psd$) &-0.957&2.362&-0.128&0.361\\
     \hline
       \end{tabular}
	\end{center}
\label{nme}
\end{table}

\section{summary and conclusion}\label{SecIV}

In the current study, we compare $0\nu\beta\beta$  NMEs predictions in light nuclei
from shell model calculations with VMC calculations.
In all cases the bare operators are used.
The VMC calculations agree well with numerically exact GMFC calculations, reproducing nuclear radii and electroweak distributions.
In this sense, the VMC calculations act as a benchmark for comparison to other models.

We study two very different double beta-decay transitions,
$^{10}$Be$\rightarrow^{10}$C and $^{12}$Be$\rightarrow^{12}$C.
The most significant difference between these two transitions is that
nuclear structure causes the $C(r)$ radial contributions
to the $0\nu\beta\beta$ matrix elements in A=10 to be nodeless,
whereas the same $C(r)$ distributions for A=12 involve nodes and hence cancellations in the  matrix elements.

We examine the role of various model-dependent choices
that go into shell model calculations of $0\nu\beta\beta$ matrix elements,
including the the size of the model space, radial wave functions,
and the SRC functions. The largest impact on the shell model predictions
comes from the many-body correlations that are introduced by increasing
the size of the model space. In both A=10 and A=12, the transition densities
$C(r)$ determining the matrix elements of $M_\alpha^\beta$ from the larger
shell model calculation come significantly closer to the predictions of the VMC
calculations than do the predictions from the smaller shell model spaces.

The change in the absolute magnitude of the $M_F$ and $M_{GT}$ $0\nu\beta\beta$ matrix
elements with increasing shell model space is different for A=10 and A=12.
In both cases the magnitude of the peaks in the $C(r)$ distribution increase. In the case of  the nodeless
A=10 $C(r)$ function, this translates into an overall increase in the absolute
magnitude of the $0\nu\beta\beta$ matrix elements. In contrast, the change in
the magnitude of the peaks, coupled with the shift in the position of the node, causes
the A=12 cancellations from contributions above and below the node of C(r) to reduce
(increase) the absolute magnitude of the $M_{GT}$ ($M_F$) matrix element by a factor of 2 (1.4).

The use of more realistic WS radial wave functions, that take the separation energies of the
transferred neutrons and protons into account, reduce the magnitudes of the NMEs because
of the reduced overlap between the neutron and proton wave functions. By using the WS
wave functions and increased model space, the long-range behavior of $C(r)$ distribution
is found to be in excellent  agreement with  the VMC predictions.

The SRC functions affect the contributions to the matrix elements at short distance (less than 1.5 fm).
We have examined several different SRC functions. In general, the introduction of
SRCs moves the shell model predictions closer to the VMC calculations, reducing the NMEs.
The magnitude of this reduction is dependent on the SRC function used.


While the present calculations in light nuclei cannot be used to make
definitive statements about the validity of shell model calculations in
medium and heavy nuclei, they do suggest some trends.
First, the use of H.O. radial wave functions will likely lead to an overestimate of matrix elements.
Second, and perhaps more importantly, limited size model space calculations could
affect the magnitude of the predicted $0\nu\beta\beta$ matrix elements,
particularly for calculations constrained to a single shell.
Third, the inclusion of a SRC function is needed. The best choice for this function requires further study.

\acknowledgments{}
We thank V. Cirigliano and W. Dekens for helpful discussions.
X.B. Wang and G.X. Dong thank the support by National Natural Science Foundation of China
under Grant Nos. U1732138, 11605054, 11505056, and 11847315, and especially thank the hospitality and financial
support of Los Alamos National Laboratory.
This research
is also supported by the U.S.~Department of Energy, Office of Science,  Office of
Nuclear Physics, under contracts
DE-AC02-06CH11357 (R.B.W.), and DE-AC52-06NA25396 and
the Los Alamos LDRD program. The work R.B.Wiringa has been supported by the Nuclear
Computational Low-Energy Initiative (NUCLEI) SciDAC project.
Computational resources have been provided by Los Alamos Open
Supercomputing, and Argonne's Laboratory Computing Resource Center.


\bibliographystyle{apsrev4-1}
\bibliography{bibliography}

\end{document}